\documentclass[adp,fleqn]{w-art}
\usepackage{times}
\usepackage{w-thm}
\usepackage[]{graphicx}
\chardef\bslash=`\\ 

\begin{document}
\DOIsuffix{theDOIsuffix}
\Volume{X} \Issue{X} \Copyrightissue{X} \Month{X} \Year{X} 
\pagespan{1}{}
\Receiveddate{dd mm yyyy} 
\Accepteddate{dd mm yyyy}



\title[Propagation of extended objects across singularity]{Propagation of extended objects across singularity of time\\ dependent orbifold}

\author[P.\ Ma\l kiewicz]{Przemys\l aw Ma\l kiewicz\footnote{E-mail: {\sf pmalk@fuw.edu.pl}, Phone: (+48 22) 55 32 275,
     Fax: (+48 22) 62 16 085}}
\address{Theoretical Physics Department,\\ Institute for Nuclear Studies, Hoza 69, Warsaw}

\begin{abstract}
  In this paper we argue that the compactified Milne space is
  a promising model of the cosmological singularity. It is shown that
  extended objects like strings propagate in a well-defined manner
  across the singularity of the embedding space. Then a proposal
  for quantization of extended objects in the case of
  a membrane is given.
\end{abstract}
\maketitle

\section{Introduction}
\label{sect1}

One of the simplest models of the neighborhood of the cosmological
singularity, inspired by string/M theory \cite{Khoury:2001bz}, is
the compactified Milne space, $\mathcal{M}_C$. It has been used in
the cyclic universe scenario
\cite{Khoury:2001bz,Steinhardt:2001vw,Steinhardt:2001st}. This
model seems to be attractive, because it consists of
pre-singularity and post-singularity epochs and can be described
in terms of {\it quantum} elementary objects propagating in {\it
classical} spacetime.

Let us consider a two-dimensional spacetime with the line element:
\begin{equation}\label{line1}
    ds^2 =  - dt^2 +t^2 d\theta^2 .
\end{equation}
We identify the points $\theta \sim \theta+\beta$ for some fixed
value of $\beta$, so that $\theta\in [0,\beta[$. Generalization of
(\ref{line1}) to the $d+1$ dimensional spacetime, which will be
denoted by $\mathcal{M}_C$, has the form:
\begin{equation}\label{line2}
ds^2 = -dt^2  +  t^2 d\theta^2 + \delta_{kl}~dx^k dx^l ,
\end{equation}
where $t,x^k \in {R},~\theta\in {S}^1~(k= 2,\ldots, d)$.

One term in the metric (\ref{line2}) disappears/appears at $t=0$,
thus the space $\mathcal{M}_C$ may be used to model the
big-crunch/big-bang type singularity. Orbifolding  ${S}^1$ to the
segment gives a model of spacetime in the form of two orbifold
planes which collide and re-emerge at $t=0$. Our results apply to
both choices  of topology of the compact dimension.

The Polyakov action integral for a test $p$-brane (i.e. $0$-brane
= particle, $1$-brane = string, $2$-brane = membrane, $\dots$)
embedded in a fixed background spacetime with metric $g_{\mu\nu}$
reads:
\begin{equation}
    S_p= -\frac{1}{2}\mu_p \int d^{p+1}\sigma
    \sqrt{-\gamma}\;\big[\gamma^{ab}\partial_a X^{\mu} \partial_b
    X^{\nu}
    g_{\mu\nu}-(p-1)]
\end{equation}
where $\mu_p~$ is  mass per unit of $p$-volume, $\sigma^a$ are
$p$-brane worldvolume coordinates, $\gamma_{ab}~$ is  $p$-brane
worldvolume metric, $\gamma := det[\gamma_{ab}]$, $(X^{\mu})\equiv
(T,X^k,\Theta)\equiv (T,X^1,\ldots,X^{d-1},\Theta)$ are  embedding
functions of $p$-brane, i.e. $X^{\mu} = X^{\mu}(\sigma^a)$,
corresponding to $(t,x^1,\ldots,x^{d-1},\theta)$ directions of
$d+1$ dimensional background spacetime. The case of a particle
propagating in $\mathcal{M}_C$ is not clear and was studied in
\cite{Malkiewicz:2005ii, Malkiewicz:2006wq}.

\section{Dynamics of classical string}

Propagation of classical string across the singularity of
$\mathcal{M}_C$ is the best example of how extended objects can
successfully 'cure' spacetime singularities. In what follows we
use the {\it local flatness} of $\mathcal{M}_C$ to solve the
dynamics of a string. The well-known string's propagation in
Minkowski space is given by:
\begin{equation}\label{mink}
  x^{\mu}(\tau,\sigma) = x^{\mu}_+(\tau+\sigma)+ x^{\mu}_-(\tau-\sigma),
 \end{equation}
\begin{equation}\label{gauge2}
  \partial_{\tau}x^{\mu}\partial_{\tau}x_{\mu}
  +\partial_{\sigma}x^{\mu}\partial_{\sigma}x_{\mu} =0,~~~~~~
\partial_{\tau}x^{\mu}\partial_{\sigma}x_{\mu}=0 ,
\end{equation}
where $x^{\mu}$ $(\mu=0,1,\dots)$ are Minkowski coordinates and
$x_\pm^\mu$ are any functions. The equations (\ref{gauge2}) are
just gauge constraints. For {\it winding} modes
$\overline{x}(t,\theta)$ in $\mathcal{M}_C$, where
$\overline{x}:=(x^2, x^3, \texttt{\ldots}, x^d)$ one shows in
\cite{Malkiewicz:2008dw} that the extra conditions hold:
\begin{equation}\label{sym}
  x^0 = f(\tau+\sigma)-f(-\tau+\sigma),~~~~~
  x^1 = g(\tau+\sigma)-g(-\tau+\sigma) ,
\end{equation}
and
\begin{equation}\label{deter}
  x^{k}_+(\sigma_+)+x^{k}_-(\sigma_-) = \sum_n a_n^k(t)\exp\big
(\imath\frac{2\pi
  n}{\beta}\theta\big),~~k>1.
\end{equation}
Satisfying the last condition is not straightforward and rests
upon the fact that the dynamics is governed by a second order
differential equation. Thus it is sufficient to satisfy the
condition (\ref{deter}) by specifying $x^k$, $\partial_t x^k$ on a
single Cauchy's line. In this way one rules out one of the
variables in (\ref{deter}) and compares functions dependent on
just a single variable. This strategy works
\cite{Malkiewicz:2008dw} and leads to the solutions:
\begin{eqnarray}\label{solution}
  x^0 &=& q\sinh(\sigma_+)+q\sinh(\sigma_-),~~x^1 = q\cosh(\sigma_+)-q\cosh(\sigma_-),\\
 \label{solution2}
  x^k &=& \sum_n a_{n+}^k \exp\big (\imath\frac{2\pi
  n}{\beta}\sigma_+\big ) +\sum_n a_{n-}^k \exp\big (\imath\frac{2\pi
  n}{\beta}\sigma_-\big)+{c_0^k}(\sigma_++\sigma_-),
\end{eqnarray}
where $k>1$ and $a_{n+}^k,~a_{n-}^k,~c_0^k$ are constants. These
solutions should satisfy the gauge conditions (\ref{gauge2}),
which in the case of $\mathcal{M}_C$ read
\begin{equation}\label{gauge}
  \partial_+x_k\partial_+x^k = q^2 =
  \partial_-x_k\partial_-x^k .
\end{equation}
Alternatively, the solutions (\ref{solution2}) in terms of $t$ and
$\theta$ have the form
\begin{eqnarray}\label{gen}
  x^k(t,\theta) &=& \sum_n \Big ( a_{n+}^k e^{\imath\frac{2\pi
  n}{\beta}\textrm{arcsinh}\big (\frac{t}{2q}\big )}+a_{n-}^k e^{-\imath\frac{2\pi
  n}{\beta}\textrm{arcsinh}\big (\frac{t}{2q}\big )}\Big )  \exp\big (\imath\frac{2\pi
  n}{\beta}\theta\big )
  \nonumber\\&+&2c_0^k\textrm{arcsinh}\Big (\frac{t}{2q}\Big ),
\end{eqnarray}
where $n$ denotes $n$-th excitation. The number of arbitrary
constants  in (\ref{gen}) can be reduced by the imposition of the
gauge condition (\ref{gauge}).

One observes that the above solutions are well-defined everywhere
and it is reasonable to expect that the same holds for higher
dimensional objects like classical membrane. The quantization
should not spoil this as it was proven in the case of a string in
\cite{Malkiewicz:2006bw}.

\section{Canonical quantization of membrane}

Total Hamiltonian, $H_T$, corresponding to the Polyakov action
reads (see e.g. \cite{Turok:2004gb}):
\begin{equation}
H_T = \int d^p\sigma \mathcal{H}_T\ ,\ \ \mathcal{H}_T := A C +
A^i C_i,
\end{equation}
where $A=A(\sigma^a)$ and $A^i= A^i(\sigma^a)$ are any `regular'
functions, and  $~C~$ and $~C_i~$ are first-class constraints:
\begin{equation}
  C = \Pi_\mu\;\Pi_\nu\;g^{\mu\nu} + {\mu}_p^2 \;\partial_\sigma X^{\mu}\partial_\sigma
  X^{\nu}g_{\mu\nu}\approx 0\ ,\ \ C_i = \Pi_\mu\partial_{i}X^{\mu}\approx
  0,~~(i=1,\dots,p)
\end{equation}
with Poisson bracket $\{\cdot,\cdot\}:= \int
d^p\sigma\Big(\frac{\partial\cdot}{\partial X^{\mu}}
\frac{\partial\cdot}{\partial \Pi_{\mu}}
     - \frac{\partial\cdot}{\partial \Pi_{\mu}}
    \frac{\partial\cdot}{\partial X^{\mu}}\Big)$.
We will consider {\it uniformly winding} modes of $p$-branes in
$\mathcal{M}_C$, i.e. $\sigma^p = \theta=\Theta$ and
$\partial_\theta X^\mu=0=\partial_\theta\Pi_\mu$. This reduces
number of world-volume coordinates and subsequently number of
constraints by one so that it is now equivalent to the dynamics of
a $(p-1)$-brane in the $d$-dimensional 'flat' FRW universe with
the metric $ds^2_{red}=T\eta_{\mu\nu}~dX^{\mu}dX^{\nu}$. In the
case of membrane it leads to two constraints of the form:
\begin{equation}\label{cons}
    C:=\frac{1}{2\mu_2\theta_0  T}\Pi_{\alpha} \Pi_{\beta}\eta^{\alpha\beta} +
    \frac{\mu_2\theta_0}{2} \;T\;\partial_a X^{\alpha} \partial_b
    X^{\beta} \eta_{\alpha\beta}\approx 0\ ,\ \ C_1 := \partial_{\sigma} X^{\alpha} \Pi_{\alpha} \approx 0,
\end{equation}
which effectively are constraints of a string in the spacetime
with the line element $ds^2_{red}$. We redefine the constraints
(\ref{cons}) and smear them with test functions:
\begin{equation}
L^{\pm}_n :=\int{C}_{\pm}(\sigma)\cdot\exp{(in\sigma)}~d\sigma
,~~n \in Z\ ,\ \ C_{\pm}:=\frac{C\pm C_1}{2}
\end{equation}
and check that the new constraints satisfy the following Lie
algebra:
\begin{equation}\label{consL}
\{L^+_n,L^+_m\} = i(m-n) L^+_{m+n}\ , \ \ \{L^-_n,L^-_m\} = i(m-n)
L^-_{m+n}\ , \ \ \{L^+_n,L^-_m\} = 0,
\end{equation}
where $(L^{\pm}_n)^{\ast}={L}^{\pm}_{-n}$. Now we define Hilbert
space encoding many-field degrees of freedom as in \cite{DQM}:
\begin{equation}
  \mathcal{H}\ni\Psi[\overrightarrow{Y}] := \int \psi(\overrightarrow{Y},
  \acute{\overrightarrow{Y}},\sigma)d\sigma
  ,~\overrightarrow{Y}:=\overrightarrow{Y}(\sigma)
\end{equation}
such that $\|\Psi
  \|<\infty$ and $\langle\Psi |\Phi \rangle := \int\overline{\Psi}[\overrightarrow{Y}]
  \Phi[\overrightarrow{Y}][d\overrightarrow{Y}]$. We define the
  operators $\hat{L}_{n}$ as follows:
\begin{eqnarray}\nonumber
\hat{L}_{n}\Psi[\overrightarrow{Y}]&:=&i\int
\Big(\frac{\partial\psi}{\partial
Y^{\mu}}e^{in\sigma}\frac{d}{d\sigma}Y^{\mu}+\frac{\partial\psi}{\partial
\acute{Y}^{\mu}}\frac{d}{d\sigma}[e^{in\sigma}\frac{d}{d\sigma}{Y^{\mu}}]\Big)d\sigma\\
&=&\int e^{in\sigma}\Big(-i\frac{\partial\psi}{\partial
\sigma}+n\frac{\partial\psi}{\partial
\acute{Y^{\mu}}}\acute{Y}^{\mu}-n\psi\Big)d\sigma~\in\mathcal{H}
\end{eqnarray}
One may check that:
\begin{equation}
[\hat{L}_n,\hat{L}_m] = (n - m) \hat{L}_{n+m} \textrm{,~~~~} \ \
\langle\hat{L}_n\Psi |\Phi \rangle=\langle\Psi
|\hat{L}_{n}^{\dag}\Phi \rangle=\langle\Psi |\hat{L}_{-n}\Phi
\rangle,
\end{equation}
which is a quantum counterpart for each subalgebra contained in
the full algebra (\ref{consL}). To construct the representation of
the full algebra (\ref{consL}), which consists of two commuting
subalgebras, one may use standard techniques, i.e. direct sum or
tensor product of the representations of both subalgebras. Now,
following the Dirac prescription one solves the quantum
constraints, i.e. one looks for such $\Psi$ that:
\begin{equation}\label{dirac}
\hat{L}_n\Psi[\overrightarrow{Y}]=0,~~n\in{Z}.
\end{equation}
For $\psi=\psi(\overrightarrow{Y},\acute{\overrightarrow{Y}})$ the
condition (\ref{dirac}) reads:
\begin{equation}
\int \acute{(e^{\imath
n\sigma})}[-\psi+\frac{\partial\psi}{\partial
\acute{Y}^{\mu}}\acute{Y}^{\mu}]~d\sigma=0,~~n\in{Z},
\end{equation}
which has the solution \cite{DQM}:
\begin{equation}\label{dirsol}
\psi=\bigg(\sum_i\alpha_i(\overrightarrow{Y})\prod_{\mu}|
\acute{Y}^{\mu}|^{\rho_i^{\mu}}\bigg)^{\frac{1}{\rho}}-c ,
\end{equation}
where $\sum_{\mu}\rho_i^{\mu}=\rho$. This is an expected result
since  the measure
$\sqrt[\rho]{\prod_{\mu}|\acute{Y}^{\mu}|^{\rho^{\mu}}}d\sigma$ is
invariant with respect to $\sigma$-diffeomorphisms.

All operators acting on the solutions (\ref{dirsol}) are
observables since they act on gauge invariant states. The whole
variety of states includes many subspaces, which we can use to
construct representations of observables. An example of such
subspace is spanned by:
\begin{equation}
\psi :=
\alpha_{\mu}(\overrightarrow{Y})\acute{Y}^{\mu}~~~~\Longrightarrow~~
\Psi[Y]=\int\alpha_{\mu}(\overrightarrow{Y})d{Y}^{\mu}.
\end{equation}
We may introduce quantum observables
$\widehat{O}_s,~s=s^{\lambda}({Y}^{\mu})\partial_{{Y}^{\lambda}}$
such that:
\begin{equation}\label{obs}
    \widehat{O}_s\bigg(\int\alpha_{\mu}d{Y}^{\mu}\bigg)=\int\big(
s^{\lambda}\alpha_{\mu,\lambda}+s^{\lambda}_{,\mu}\alpha_{\lambda}\big)d{Y}^{\mu}~~\Longrightarrow~~[\widehat{O}_s,\widehat{O}_t
]=\widehat{O}_{[s,t]}.
\end{equation}

But what are the fields $Y^{\mu}$? It seems that one needs to
postulate (find?) a relation between $Y^{\mu}$ and
$\{{X}^{\mu}\times\acute{X}^{\mu}\times\Pi_{\nu}\}$. Such a
relation was proposed in \cite{DQM}. Study of this relation would
enable to interpret the observables in (\ref{obs}) in physical
terms and thus complete the proposed quantization scheme for
membrane.

\section{Conclusions}

It seems that the compactified Milne space, $\mathcal{M}_C$, is
suitable for modelling higher dimensional cosmological
singularity. We showed that classical propagation of excited
string is well-defined and unambiguous. The natural expectation
would be: quantization should not spoil it!

We have proposed a quantization procedure for uniformly winding
membrane, within which we made some progress, particularly we
found non-trivial quantum states. It would be interesting to find
some relation of our quantization of membrane with M-theory (in
our procedure there is no critical dimensionality). However, our
work is a first step toward the full resolution of the
cosmological singularity that would require quantization of both
spacetime and physical $p$-branes.

\begin{acknowledgement}
This work has been supported by the Polish Ministry of Science and
Higher Education Grant  NN 202 0542 33.
\end{acknowledgement}

\end{document}